\documentclass[reprint,superscriptaddress]{revtex4}
\usepackage{bm}
\usepackage{amssymb}
\usepackage{amsfonts}
\usepackage{graphicx}
\usepackage{subfigure}
\usepackage{multirow}

\begin{document}

\title{Two-Qubit Gate Operation on Selected Nearest-Neighbor Neutral Atom Qubits
}

\author{Elham Hosseini Lapasar}
\email{hosseini@sci.osaka-cu.ac.jp}
\affiliation{Department of Chemistry and Materials Science, Graduate School of Science, Osaka City University, Sumiyoshi, Osaka 558-8585, Japan}
\author{Kenichi Kasamatsu}
\affiliation{Department of Physics, Kinki University, Higashiosaka, Osaka 577-8502, Japan}
\affiliation{Research Center for Quantum Computing,
Interdisciplinary Graduate School of Science and Engineering, Kinki University, Higashiosaka, Osaka 577-8502, Japan}
\author{S\'{i}le Nic Chormaic}
\affiliation{Light-Matter Interactions Unit, OIST Graduate University, Onna-son, Okinawa 904-0495, Japan}
\author{Takeji Takui}
\affiliation{Department of Chemistry and Materials Science, Graduate School of Science, Osaka City University, Sumiyoshi, Osaka 558-8585, Japan}
\author{Yasushi Kondo}
\affiliation{Department of Physics, Kinki University, Higashiosaka, Osaka 577-8502, Japan}
\affiliation{Research Center for Quantum Computing,
Interdisciplinary Graduate School of Science and Engineering, Kinki University, Higashiosaka, Osaka 577-8502, Japan}
\author{Mikio Nakahara}
\affiliation{Department of Physics, Kinki University, Higashiosaka, Osaka 577-8502, Japan}
\affiliation{Research Center for Quantum Computing,
Interdisciplinary Graduate School of Science and Engineering, Kinki University, Higashiosaka, Osaka 577-8502, Japan}
\author{Tetsuo Ohmi}
\affiliation{Research Center for Quantum Computing,
Interdisciplinary Graduate School of Science and Engineering, Kinki University, Higashiosaka, Osaka 577-8502, Japan}

\begin{abstract}
We have previously discussed the design of a neutral atom quantum computer with an on-demand
 interaction [E. Hosseini Lapasar, {\it et al.}, J. Phys. Soc. Jpn. \textbf{80}, 114003 (2011)].
In this contribution, we propose an experimental method to demonstrate a selective two-qubit gate operation that is less demanding than our original proposal, although the gate operation is limited to act between two neighboring atoms. We evaluate numerically the process of a two-qubit gate operation that is applied to a selected pair of nearest-neighbor, trapped atoms and we estimate the upper bound of the gate operation time and corresponding gate fidelity. The proposed scheme is scalable and, though challenging, is feasible with current experimental capabilities.
\end{abstract}


\maketitle

\section{Introduction}
Quantum information science has rapidly grown with the promise of realizing a quantum computer that will be capable of solving many classically intractable problems by making use of properties of quantum systems, such as superposition and entanglement \cite{Nielsen,NakaharaOhmi,NatureRev}.
Similar to classical computers, a quantum computer consists of a memory and a processor, with a set of gates to store and process the information.
A number of physical systems have been proposed for practical implementations of quantum computing, such as nuclear magnetic resonance, linear optics, trapped atoms or ions, and superconducting circuits \cite{NakaharaOhmi,NatureRev}. These physical systems have merits and demerits and, to date, no single system satisfies all the DiVincenzo criteria - the necessary conditions for a physical system to be a realistic candidate for a working quantum computer \cite{DiVincenzo}.
A major obstacle in the feasibility of quantum computing is decoherence, which arises from the interaction of the system with its environment.

A system of trapped, neutral atoms is one of the most promising candidates for implementing a scalable quantum computer, since neutral atoms have the advantage of an intrinsically weak interaction with the environment \cite{Deutsch,Negretti}.
Coherence times in this system are a few seconds \cite{Treutlein}, i.e.,  many orders of magnitude longer than
typical operation times.
A one-qubit gate operation has already been demonstrated using microwave radiation \cite{Schrader} or a two-photon Raman transition \cite{Yavuz,Jones}, which is a well-established technique by today.
However, a two-qubit gate operation - and the relevant high-fidelity control with scalable architectures - is still
 a serious challenge in a neutral atom-based system.
Two-qubit gate operations have been proposed using either short-range collisions \cite{Jaksch,Calarco,Calarco2} or long-range dipole-dipole interactions \cite{Brennen,Jaksch2}; the latter scheme has been realized with individual atoms in separate dipole traps \cite{Wilk,Isenhower}.
Mandel {\it et al.} have demonstrated a two-qubit gate operation with neutral atoms using hyperfine state dependent optical lattice potentials \cite{Mandel1,Mandel2}. A drawback of this implementation is that the gate acts on all the nearest neighbor pairs in the optical lattice simultaneously and cannot be used for circuit model quantum computation.

We recently proposed a method to apply a two-qubit gate operation on an arbitrarily selected pair of atoms by modifying
 the method demonstrated by Mandel {\it et al.} \cite{Hosseini}, where the operation is implemented by
optical potentials generated by near-field Fresnel diffraction (NFFD)
of light at a thin aperture \cite{Bandi} and a state-dependent optical lattice \cite{Mandel1}.
In order to verify the feasibility of this scheme, the compatibility of the proposed two-qubit
operation \cite{Hosseini} with the NFFD trap would need to be verified.
The proposed experiment is not overly demanding, though may not be suitable for future quantum computer realizations. A similar, but alternative, system has also been proposed, based on  trapping of atoms above an array of microtraps \cite{Cirac}.

Here, we present a proposal for an experiment towards two-qubit gate operation using neutral atoms.
Although the proposed system is less general than our original proposal \cite{Hosseini}, it should be easier to implement.
We first briefly review the design of the neutral atom quantum computer proposed
in Ref. \cite{Hosseini}, and highlight  the difficulties associated with the system.
Next, we propose a demonstration of a two-qubit operation under
simplified situations obtained by avoiding these difficulties. We will finally present, in detail, numerical calculations
corresponding to the process of the two-qubit gate operation, from which we estimate the upper bound
 of the gate operation time and corresponding gate fidelity.

\section{Brief Review of Our Recent Proposal}
\begin{figure}[h]
\begin{center}
{\includegraphics[scale=0.5,clip]{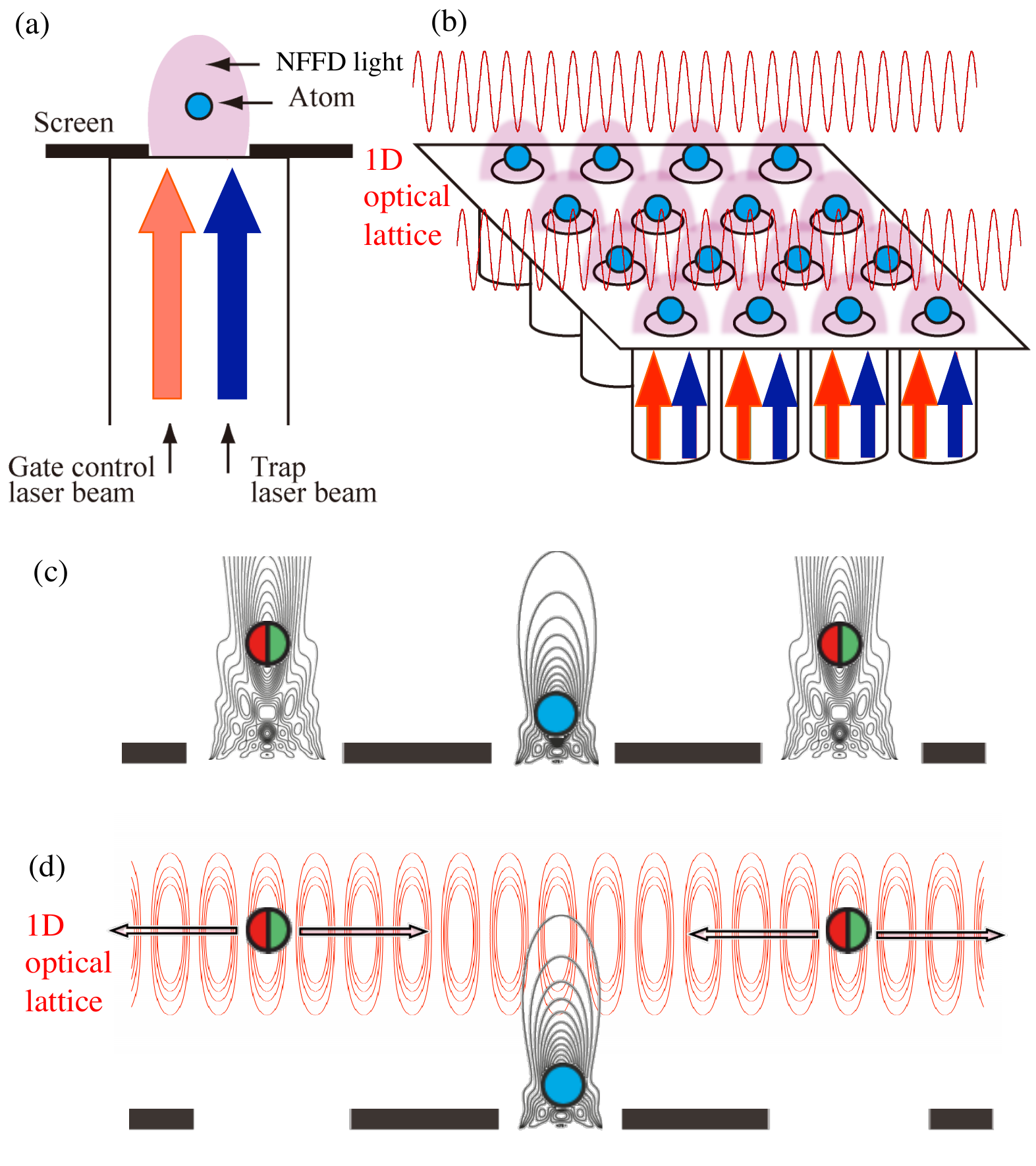}}
\caption{\label{fig:proposal0}
(Color online) Schematic illustration of our previous proposal for a neutral atom quantum computer with an on-demand interaction \cite{Hosseini}.  (a) An atom trapped in NFFD light. The blue (darker) arrow shows the trap light while the red (lighter) arrow shows the laser light required for one-qubit gate operations, in which an atom is acted upon by the Walsh-Hadamard gate in advance so that it is in a superposition of
$| 0 \rangle$ (red (darker) semicircle) and $| 1 \rangle$ (green (lighter) semicircle). (b) Example of an array of such traps. The trap laser and the gate control laser are explicitly shown only for the first row.
A one-dimensional optical lattice is superimposed near the array of NFFD traps. The minima of the optical lattice are situated so as to adjust the period of the array of the NFFD traps.
(c) Side view of the array (b).
The black contours show the individual NFFD potential given by Eq. (\ref{eq:rs}) and the bold black segments show the substrate.
By enlarging the aperture size, the minima of the selected traps are moved away from the substrate.
(d) A one-dimensional optical lattice, depicted by the red contours,
is superimposed while the selected NFFD traps are turned off adiabatically.
Two atoms in the superposition of $| 0 \rangle$ and $| 1 \rangle$ are
 left in the state-dependent optical lattice. }
\end{center}
\end{figure}

Figure \ref{fig:proposal0} shows the scheme of our recent proposal \cite{Hosseini}.
Neutral atoms with two hyperfine states, $|0\rangle$ and $|1\rangle$, are trapped
 in an array of optical  potentials generated by NFFD of light at  thin apertures, each
with a radius comparable to the wavelength of the light \cite{Bandi}.
The apertures are made using microfabrication techniques on a silicon substrate.
Two-colored laser light is  incident on the apertures through a tapered optical fiber attached
to each aperture.
One laser color is used for trapping the atom and the other is used to control the hyperfine qubit states
 of the atom, i.e., it provides a one-qubit gate operation to prepare a superposition state $(|0\rangle+|1\rangle)/\sqrt 2$,
by using the two-photon Raman transition \cite{Yavuz,Jones}.
Since each site is equipped with its own gate control laser beam, individual, single-qubit gate operations can be applied on many qubits simultaneously.
We assume that the NFFD potential is adjusted so that only one atom is contained within each trap.

In order to implement a two-qubit gate operation between an arbitrary pair of atoms,
a one-dimensional optical lattice is superimposed near the substrate
 by applying a pair of tightly-focused, counterpropagating laser beams along the atom trapping sites,
so that the minima of the optical lattice are situated at the space coordinates for the atoms trapped in NFFD potentials.
By manipulating the aperture size, one can control the position of an individual atom
 in a direction perpendicular to the aperture plane to transfer atoms into a one-dimensional optical lattice \cite{Hosseini,Nakahara,Bose},
as shown in Fig. \ref{fig:proposal0}(c).
A collision in the subspace $|0\rangle|1\rangle$ can be caused by controlling the polarization of the counterpropagating
 beams of the optical lattice [Fig. \ref{fig:proposal0}(d)], so that this particular component acquires an extra dynamical phase \cite{Jaksch,Calarco}.
Subsequently, the polarizations are reversed and the atoms are returned to their initial positions in the optical lattice.
Finally, the atoms are transferred from the optical lattice to the initial NFFD traps.
As an example, we considered two hyperfine states $|0\rangle=|F=1, m_{\rm F}=1\rangle$
and $|1\rangle= |F=2, m_{\rm F}=1\rangle$ of $^{87} {\rm Rb}$ atoms to be compatible with this scheme
 and estimated that the time required for the entangling gate operation is on the order of $10~ {\rm ms}$ with a corresponding fidelity of 0.886 \cite{Hosseini}.

\section{Two-Qubit Gate Implementation}
\begin{figure}[h]
\begin{center}
{\includegraphics[scale=0.5,clip]{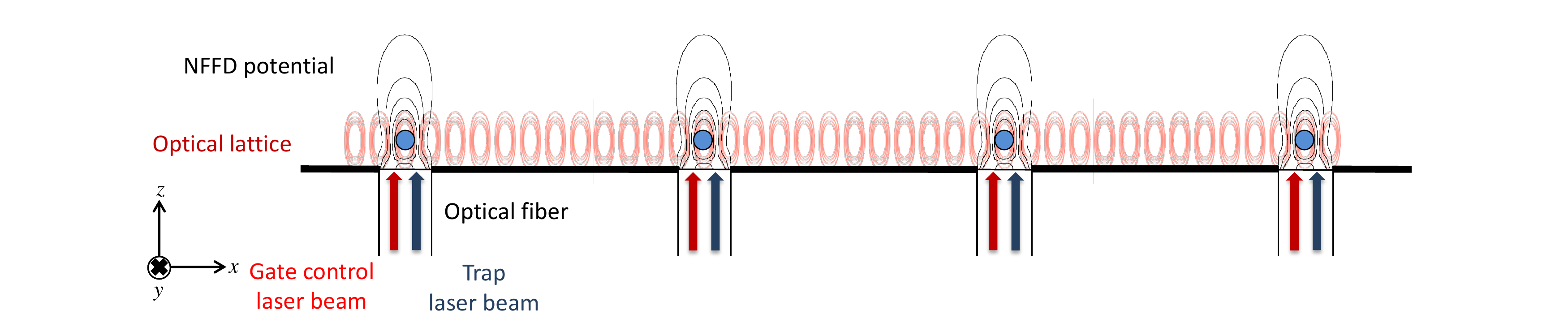}}
\caption{\label{fig:proposal}
(Color online) Schematic representation of atoms trapped in an optical lattice with an NFFD potential.
Thick and thin contours show the profiles of NFFD traps and an optical lattice, respectively.
The blue (darker) arrow shows the trapping laser beam, while the red (lighter) arrow shows the control laser beam.
Minima of the one-dimensional optical lattice along the $x$-axis are situated at the space coordinates of atoms trapped by NFFD. }
\end{center}
\end{figure}

\begin{figure}[h]
\begin{center}
{\includegraphics[scale=0.4,clip]{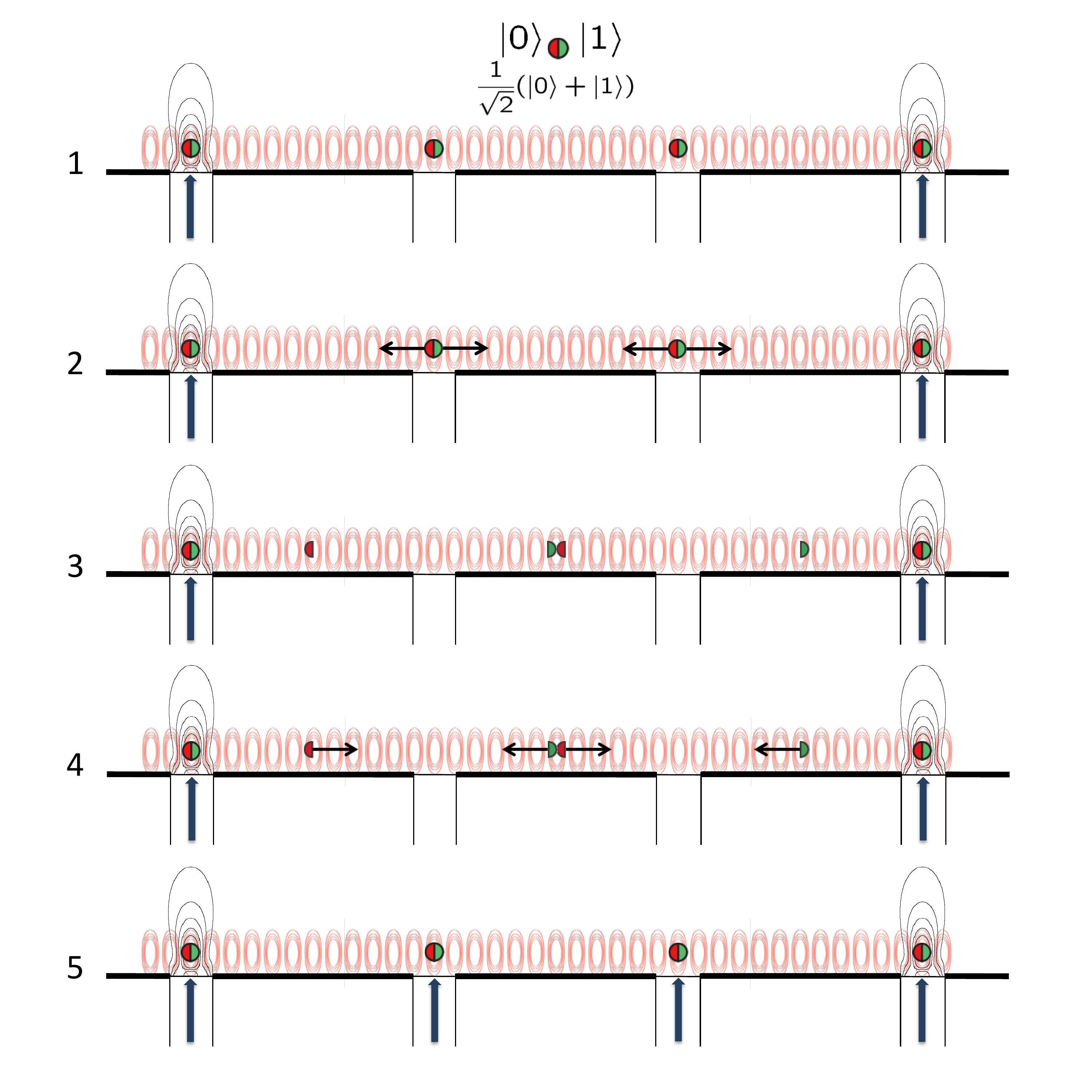}}
\caption{\label{fig:gate}
(Color online) Schematic diagram of a two-qubit gate operation applied on addressed nearest neighbor atoms.
 The Walsh-Hadamard gate is operated on all atoms in advance so that each atom is in a superposition of $|0\rangle$ and $|1\rangle$.
Thick and thin contours show the profiles of an NFFD trap potential and an optical lattice, respectively. `Spectator' atoms should be left in the ground state of the combined potential made by NFFD of light and an optical lattice during Step 2.}
\end{center}
\end{figure}

The purpose of this work is to present a quantum computer scheme that is less demanding
 than our previous proposal \cite{Hosseini}.
There are primarily three difficulties in our previous proposal:
\begin{itemize}
\item[(i)] making apertures of variable size,
\item[(ii)] aligning optical fibers on individual apertures with a size of a few micrometers,
\item[(iii)] making an optical lattice very close to the surface of the array.
\end{itemize}
The most challenging part is the aperture size control within microseconds.
To circumvent this difficulty (i), we propose an alternative design as shown in Fig. \ref{fig:proposal}.
The design is very similar to the previous one [Fig.\ref{fig:proposal0}], but the optical lattice is implemented
so that its minima are situated at the space coordinates of NFFD trapped atoms {\it from the outset}.
To implement a two-qubit gate operation, we decompose the procedure into five steps as shown in Fig. \ref{fig:gate} and outlined in the following.

\begin{enumerate}
\item
We choose two neighboring trapped atoms to be operated on by the gate.
The NFFD laser beams of the two atoms are adiabatically turned off
 so that the two atoms remain in the optical lattice.
\item
The polarizations of the pair of counterpropagating laser beams, with which the optical lattice is created, are rotated in opposite directions so that the qubit state $|0\rangle$ is transferred toward the negative $x$-direction, while $|1\rangle$ is transferred toward the positive $x$-direction \cite{Mandel1}. Therefore, it is possible to collide $|0\rangle$ of one atom and $|1\rangle$ of the other atom.
 It is important to ensure that `spectator' atoms other than the chosen atomic pair should be left
 in the ground state of the potential made by the combined NFFD and optical lattice during this process.
\item
Now, component $|0\rangle$ of one atom and $|1\rangle$ of the other atom are trapped in the same potential well in the optical lattice. They interact with each other for a duration $t_{\rm hold}$ so that the particular state $|0\rangle|1\rangle$ acquires an extra phase compared to other components $|0\rangle|0\rangle$, $|1\rangle|0\rangle$, and $|1\rangle|1\rangle$. Here, $U_{\rm int}=(4 \pi \hbar^2a_s/m)\int |\psi|^4 d{\bm r}$ is the on-site interaction energy with the atomic mass $m$ and the $s$-wave scattering length $a_s$. When $t_{\rm hold}$ satisfies the condition $U_{\rm int}t_{\rm hold}=\pi$, we obtain a nonlocal two-qubit gate $|00\rangle\langle11|-|01\rangle\langle01|+|10\rangle\langle10|+|11\rangle\langle11|$.
\item
After acquiring the phase, the two states are separated along the optical lattice by the inverse process of Step 2.
\item
The NFFD laser beams are adiabatically turned on.
\end{enumerate}

Difficulty (i) is circumvented in this new design, thereby making the method more adaptable
for experimentalists, although in the present scheme the two-qubit gate operation can only be applied
on selected nearest-neighbor trapped atoms.
We should mention that one drawback of this proposal is that we need SWAP
 gates to apply two-qubit gates between two remote atoms.
 In relation to difficulty (ii), it is by now possible to fabricate high transmission, tapered optical micro- and nanofibers relatively routinely and they have found applications in several areas of cold atom physics \cite{Morrissey}.  A half-taper, as required in the proposed setup, can be made by pulling a tapered fiber until breaking point.
To place the fiber tips close to the substrate, they would need to be guided through place
holders since they are very fragile and sensitive to vibrations.
In vacuum, the latter point is less of an issue since air currents are minimal.
We should also mention that making an optical lattice very close to the surface of the array limits the size of the 2D lattice of trapped atoms. Rayleigh length, which is distance along the propagation direction of the laser beam from the most focused point to the point at which the beam size becomes $\sqrt 2$ times larger, is
$x_{R}=\frac{\pi w^2}{\lambda_{{\rm OL}}}\simeq 40 ~{\rm \mu m}$. The focused light beam will spread, and the lower half of the beam will touch the substrate. Thus, effective range that can be used as qubit operation is $2 x_R$, i.e., $80 ~{\rm \mu m}$. Then, we can have the order of $10$ qubits in the array. It can be a small number of qubits to realize a quantum computer, but might still be good starting point to test our scheme.

\section{Numerical Calculation}
In the following, as an example, we take two hyperfine states $|0\rangle=|F=1, m_{\rm F}=1\rangle$ and $|1\rangle= |F=2, m_{\rm F}=1\rangle$ of $^{87}{\rm Rb}$ atoms and give detailed numerical evaluations corresponding to the above-mentioned processes.
We analyze each step of the two-qubit gate operation using the time-dependent Schr\"{o}dinger equation
\begin{equation}
i \hbar \frac{\partial \psi}{\partial t}=
-\frac{\hbar^2}{2 m} \nabla^2 \psi+V_{\rm ext}({\bm r},t) \psi
\label{Seq0}
\end{equation}
with experimentally realistic parameters. We estimate the time required for each step to attain a fidelity of 0.99, where the fidelity is defined as the overlap between the ground state with the final potential and time-evolved wave function, which is certainly different from the errors occurring in the Hilbert space spanned by the hyperfine states $|0\rangle$ and $|1\rangle$ to which the quantum information is encoded. Since the atomic wave function to be corrected here is not the carrier of the quantum information, the ordinary quantum error correction is not applicable to compensate for the error \cite{Nielsen}. The fidelity may be improved by spending more time for each step, which leads to longer execution time though. The lower bound of the execution time for adiabatic transport of an atom is limited by the energy difference between the ground state and the first excited state. We might also use nonadiabatic transport of an atom with shorter execution time compared to adiabatic transfer and improve the fidelity. This is realized by employing the Lewies-Riesenefeld invariant \cite{ref:lr} associated with the Hamiltonian for example \cite{ref:chen1,ref:chen2}. Application of nonadiabatic atom transport to the current problem will discuss in other paper.
Next, we give an estimate of the execution time and the fidelity of the two-qubit gate operation which is limited by the adiabatic condition.

First, we give the form of the trapping potential and optical lattice used in our calculations. Parameters are given in Table \ref{tab:1}.
A red-detuned laser beam passing through an aperture with a radius on the order of the laser wavelength forms an attractive potential which is given by \cite{Bandi}
\begin{equation}
U_{\rm F}({\bm r}) = -U_0 \frac{|\mathcal{E}({\bm r})|^2}{E_0^2} \label{eq:rs},
\end{equation}
where
\begin{equation}
\mathcal{E}({\bm r})= \frac{E_0}{2\pi} \int \int
\frac{e^{ik_{\rm F} r}}{r} \frac{z}{r}
\left(\frac{1}{r}-ik_{\rm F} \right) dx'dy',
\end{equation}
and
\begin{equation}
U_0 = \frac{3}{8} \frac{\Gamma_e}{|\Delta_{eg}|} \frac{E_0^2}{k_{\rm F}^3},
\end{equation}
with the distance $r$ between ${\bm r}=(x,y,z)$
and ${\bm r}'=(x', y', 0)$ in the plane.
Here, $\Gamma_e$ is half of the spontaneous decay rate,
$E_0$ and $k_{\rm F}=2 \pi/\lambda_{{\rm F}}$ are the amplitude and wave number of the incident plane wave to the aperture from the $-z$-axis.
We use the $D_1$ transition of the $^{87} {\rm Rb}$ atom for trapping, so the detuning is
\begin{equation}
\Delta_{eg}=\omega_{\rm L}-\omega_0=2\pi c\left(\frac{1}{\lambda_{\rm F}}-\frac{1}{\lambda_0}\right)=-4.1\times 10^{11}~{\rm Hz}.
\end{equation}
which is negative for red detuning. $U_0$ is the potential depth at $t=0$; we have to choose its value so that the states of trapped spectator atoms are unchanged while the two-qubit gate operation is implemented. The integral is over the aperture region $(x'^2+y'^2 \leq a^2)$.
The NFFD trap laser intensity $I_0$ is related to the amplitude of the electric field $E_0$ as $I_0 = c E_0^2/8 \pi$. The power of each NFFD trap can be obtained by multiplying
$I_0$ and the area of the aperture, that is expected to be approximately $1.1\times 10^{-2}~{\rm W}$.
The optical lattice potential, which is made by a pair of counterpropagating laser beams with the same frequency, amplitude, and polarization along the $x$-axis, is introduced as
\begin{equation}
V_{{\rm OL}}({\bm r}) = -V_0 \cos^2(k_{{\rm OL}}x) e^{{-2 (y^2+(z-z_{\rm m})^2)/w^2}}.
\end{equation}
Here, $k_{{\rm OL}}$ and $\lambda_{{\rm OL}}$ are the wave number and wavelength of the lasers, $w=4\lambda_{{\rm OL}}$ the waist size, $z_{\rm m}$ the $z$-coordinate corresponding to the minimum of the NFFD potential, and
$V_0$ the potential depth at $t=0$. The lattice constant is always $\lambda_{{\rm OL}}/2$, whose value multiplied by integers
 should be matched to the separation between the neighboring apertures.

\begin{table}[h]
\caption{Parameters used in calculations.}
\centering
\label{tab:1}
\begin{tabular}{c c c}
\hline\hline
Radius of an aperture ~~~ & $a=1.5~\lambda_{\rm F}$ ~~~ & $\simeq 1.2~{\rm \mu m}$\\

Trap laser beam wavelength ~~~ & $\lambda_{\rm F}$ ~~~ & $795.118~{\rm nm}$\\

Trap laser intensity ~~~ & $I_{\rm 0}$ ~~~ & $\simeq 2.5\times 10^5~{\rm W/cm^2}$\\

$D_1$ transition of $^{87} {\rm Rb}$ ~~~ & $\lambda_0$ ~~~ & $794.979~{\rm nm}$\\

Minimum of NFFD potential ~~~ & $z_{\rm m}$ ~~~ & $\simeq 1.7~{\rm \mu m}$\\

Optical lattice laser beam  wavelength~~~ & $\lambda_{\rm OL}$ ~~~ & $785~{\rm nm}$\\

Depth of trapping potential at $t=0$ ~~~ & $U_0$ ~~~ & $h\times 1.03\times 10^{6}~{\rm Hz}$\\

Depth of optical lattice potential at $t=0$ ~~~ & $V_0$ ~~~ & $h\times 1.47\times 10^{5}~{\rm Hz}$\\

\hline
\end{tabular}
\end{table}

\subsection{Step 1}
\begin{figure}[h]
\begin{center}
{\includegraphics[scale=0.7,clip]{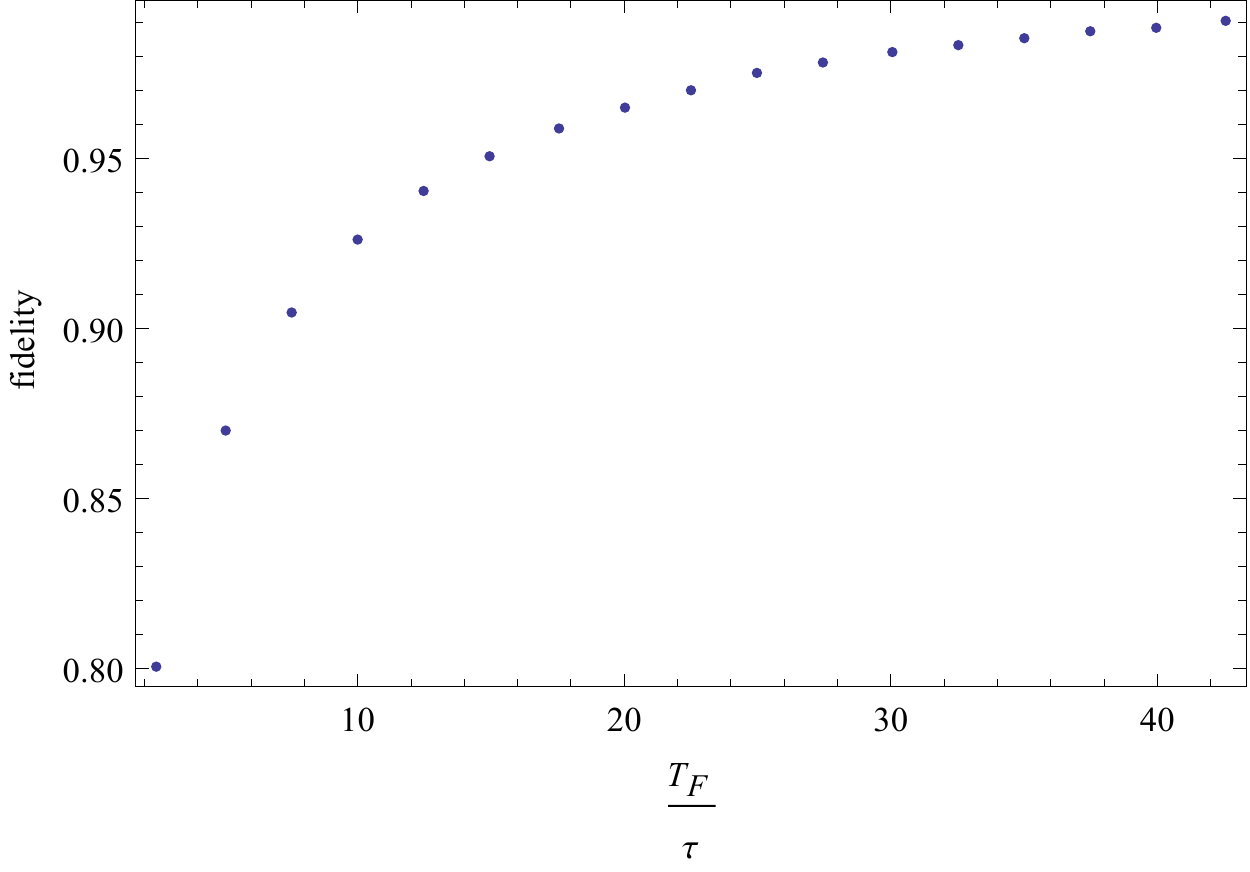}}
\caption{\label{fig:fidelity1}
(Color online) Fidelity as a function of dimensionless operation time when the NFFD potential is adiabatically turned off.
Fidelity of 0.99 is attained when $T_{{\rm F}}/\tau=42.5$.
}
\end{center}
\end{figure}

In this step we want to leave two neighboring atoms in an optical lattice to apply the gate operation on them. Therefore, we should turn off the NFFD trap potential gradually so that the atoms' wave functions are transformed adiabatically from the ground state of the combined potentials $V_{\rm OL} + U_{\rm F}$ to that of the optical lattice only. The NFFD potential is switched off as
\begin{equation}
U_{\rm F}({\bm r},t)=\cos^2\left(\frac{\pi}{2T_{{\rm F}}}t\right) U_{\rm F}({\bm r})
\end{equation}
with the operation time $T_{F}$.
To estimate the time required for this process, we need to solve numerically the
time-dependent Schr\"odinger equation Eq. (\ref{Seq0}) with
$V_{\rm ext}({\bm r},t) = V_{\rm OL} ({\bm r},t) + U_{\rm F} ({\bm r},t)$.
We have taken $V_0/E_{\rm r} = 40, U_0/E_{\rm r}=280$ and $w/\lambda_{\rm OL}=~4$ with
the scale of length $\lambda_{\rm OL} = 785{\rm nm}$,
energy $E_{\rm r}=\hbar^2 k_{\rm OL}^2/2m = h\times3.7\times10^{3}~{\rm Hz}$,
and time $\tau=\hbar/E_{\rm r} = 43~{\rm \mu s}$.
Figure \ref{fig:fidelity1} shows the fidelity as a function of dimensionless time $T_{\rm F}/\tau$,
which indicates that a time $T_{\rm F}= 1.8~{\rm ms}$ is required to attain a fidelity of $0.99$.

\subsection{Step 2}
Let us consider two atoms in the optical lattice on which the gate acts. We must put the $|0 \rangle$ state of one atom and $|1\rangle$ of the other atom in a single potential well so that the component $|0 \rangle |1 \rangle$ acquires an extra phase factor.

Suppose the polarizations of laser beams propagating along the $x$-axis and $-x$-axis are tilted by angles $\theta$ and $-\theta$ from the $z$-axis, respectively. The time-dependent polarization in the counterpropagating laser beams is introduced as
\begin{equation}
\begin{array}{c}
\bm{E}_+({\bm r}) \propto e^{ik_{\rm OL} x}(\hat{\bm{z}} \cos \theta +
\hat{\bm{y}} \sin \theta),
\vspace{.2cm}\\
\bm{E}_-({\bm r}) \propto e^{-ik_{\rm OL} x}(\hat{\bm{z}} \cos \theta -
\hat{\bm{y}} \sin \theta).
\end{array}
\end{equation}
Superposition of these counterpropagating laser beams produces an optical potential of the form
\begin{equation}\label{eq:pm}
\bm{E}_+({\bm r})+\bm{E}_-({\bm r}) \propto \sigma^+ \cos (k_{\rm OL}x-\theta)
+ \sigma^- \cos (k_{\rm OL}x+\theta),
\end{equation}
where $\sigma^+$ ($\sigma^-$) denotes a counterclockwise (clockwise) circular polarization vector. This means that the first component of Eq.~(\ref{eq:pm}) moves along the $x$-axis as $\theta$ is increased, while the second component moves along the $-x$-axis under this change. The component $\sigma^+$ introduces transitions between
fine structures; $nS_{1/2}\ (m_J=-1/2) \to nP_{1/2}\ (m_J=1/2)$, $nS_{1/2}\ (m_J=-1/2) \to nP_{3/2}\ (m_J= 1/2)$, and $nS_{1/2}\ (m_J= 1/2) \to nP_{3/2}\ (m_J=3/2)$. The transitions from $nS_{1/2}$ to $nP_{3/2}$ are red-detuned, while the transition from $nS_{1/2}$ to $nP_{1/2}$ is blue-detuned if $\omega_{\rm L}$ is chosen between the transition frequencies of $nS_{1/2}\ (m_J=-1/2) \to nP_{3/2}\ (m_J=1/2)$ and $nS_{1/2}\ (m_J=-1/2) \to nP_{1/2}\ (m_J=1/2)$. Then by adjusting $\omega_{\rm L}$ properly, it is possible to cancel the attractive potential and the repulsive potential associated with these transitions. The net contribution of the $\sigma^+$ laser beam, in this case, is an attractive
potential $V_+({\bm r}) \propto \cos^2 (k_{\rm OL}x - \theta) $ for an atom in the state $nS_{1/2}\ (m_J= 1/2)$. Similarly, the $\sigma^-$ component introduces a net attractive potential $V_-({\bm r})  \propto \cos^2 (k_{\rm OL}x + \theta)$, through the transition $nS_{1/2}\ (m_J=-1/2) \to nP_{3/2}\ (m_J=-3/2)$ on an atom in the state $nS_{1/2}\ (m_J=-1/2)$.

In summary, optical potentials for the states $S_{1/2}$ with $m_J=\pm 1/2$ are introduced as
\begin{equation}
V_{\pm} ({\bm r}) = - V_{0} \cos^2 (k_{\rm OL} x\mp \theta) e^{{-2 (y^2+(z-z_{\rm m})^2)/w^2}}
\end{equation}
and the effective potentials acting on $|0 \rangle$ and $|1 \rangle$ are evaluated as
\begin{equation}
\begin{array}{c}
\displaystyle  V_{|0 \rangle}({\bm r}) = \frac{1}{4}V_+({\bm r}) + \frac{3}{4}V_-({\bm r}),
\vspace{.2cm}\\
\displaystyle  V_{|1 \rangle}({\bm r}) = \frac{3}{4}V_+({\bm r}) + \frac{1}{4}V_-({\bm r}),
\end{array}
\end{equation}
which have been obtained by making use of the decompositions
\begin{equation}
\begin{array}{c}
\displaystyle |0 \rangle =
|F =1, m_F=1 \rangle
= -\frac{1}{2}\left|\frac{3}{2}, \frac{1}{2} \right\rangle  \left|\frac{1}{2},
\frac{1}{2} \right\rangle +
\frac{\sqrt{3}}{2} \left|\frac{3}{2}, \frac{3}{2} \right \rangle
\left| \frac{1}{2}, -\frac{1}{2} \right\rangle,
\vspace{.2cm}\\
\displaystyle  |1 \rangle =
|F= 2, m_F = 1 \rangle
=\frac{\sqrt{3}}{2}\left|\frac{3}{2},\frac{1}{2} \right\rangle
\left|\frac{1}{2}, \frac{1}{2} \right\rangle+
\frac{1}{2}\left|\frac{3}{2}, \frac{3}{2}
\right\rangle
\left|\frac{1}{2}, -\frac{1}{2}\right\rangle,
\end{array}
\end{equation}
where $|3/2, 1/2 \rangle|1/2,1/2 \rangle$ in the right hand side denotes a vector with nuclear spin $(I = 3/2, I_z=1/2)$ and electron spin $(S=1/2, S_z=1/2)$, for example. We note that $V_{|0 \rangle}(x)$ ($V_{|1 \rangle}(x)$) moves left (right) if $\theta$ is increased, which implies that the components $|0 \rangle$ and $|1 \rangle$ move in opposite directions as $\theta$ is changed. Therefore, it is possible to collide $|0 \rangle$ of one atom and $|1 \rangle$ of the other atom by changing $\theta$.

\begin{figure}[h]
\begin{center}
{\includegraphics[scale=0.7,clip]{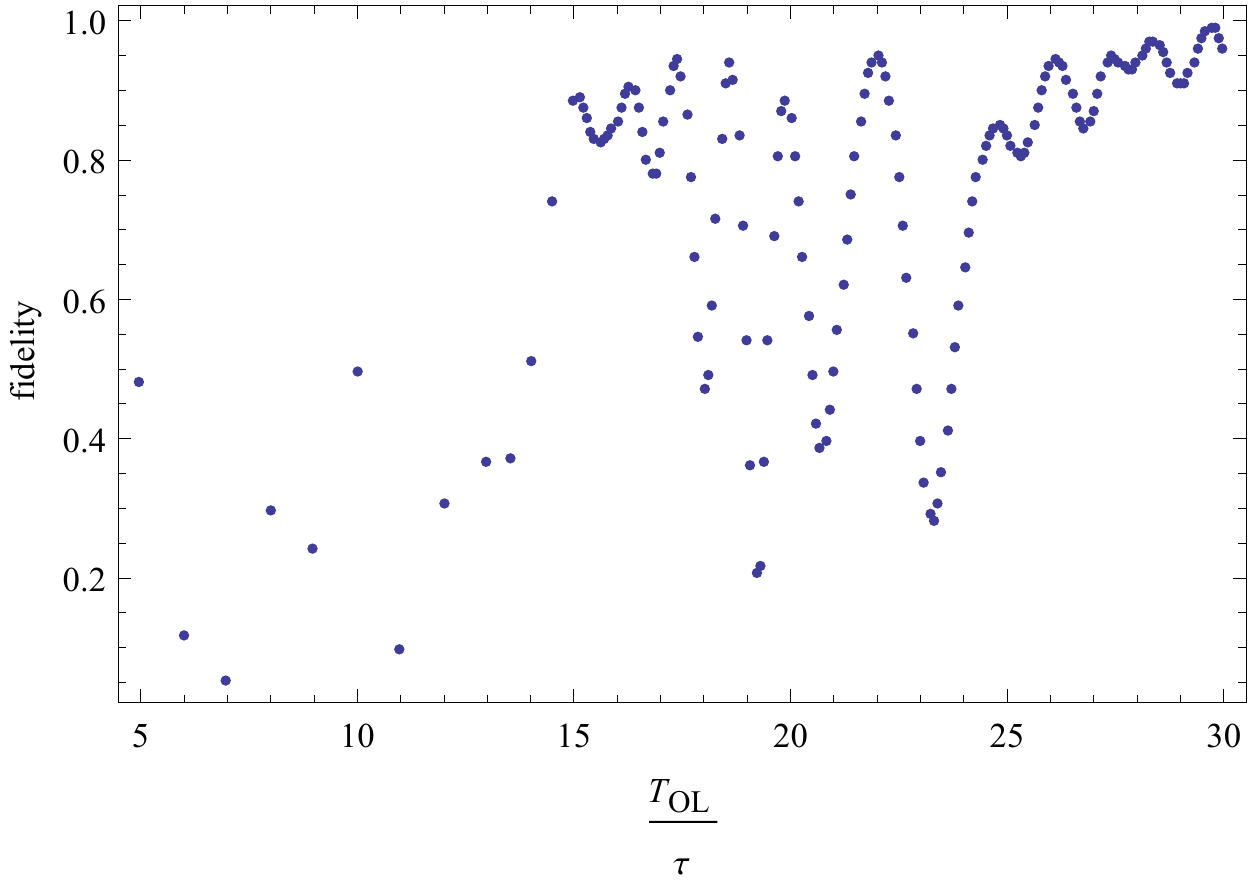}}
\caption{\label{fig:fidelity2}
(Color online) Fidelity as a function of dimensionless time when the state $|1 \rangle$ of an atom is transferred in the optical lattice by six wavelengths.
Fidelity of 0.99 is attained when $T_{{\rm OL}}/\tau=29.7$.}
\end{center}
\end{figure}

We consider motion of the $|1 \rangle$-state and solve the Schr\"odinger equation Eq. (\ref{Seq0}) with
 the potential $V_{\rm ext} = V_{|1 \rangle}({\bm r},t)$.
The angle $\theta$ is changed as
\begin{equation}
\theta(t) = n \pi \sin^2 \left(\frac{\pi }{2T_{\rm OL} } t
\right),
\end{equation}
in which a state of an atom is transferred in the optical lattice by $n$ wavelengths.
In this calculation we have taken $n=6$, which may be a minimal value for our design based on the comparison with the aperture size.
Figure \ref{fig:fidelity2} shows the fidelity as a function of dimensionless time $T_{{\rm OL}}/\tau$ for $n=6$. The fidelity is an oscillating function due to the fact that the lattice potentials $V_{|0 \rangle}$ and $V_{|1\rangle}$ are superpositions of two counterpropagating plane waves $V_{\pm}$ and there is  interference between them. We have found that the operation time $T_{\rm OL}=1.28~{\rm ms}$, yielding a fidelity of 0.99.

\begin{figure}[h]
\begin{center}
{\includegraphics[scale=0.7,clip]{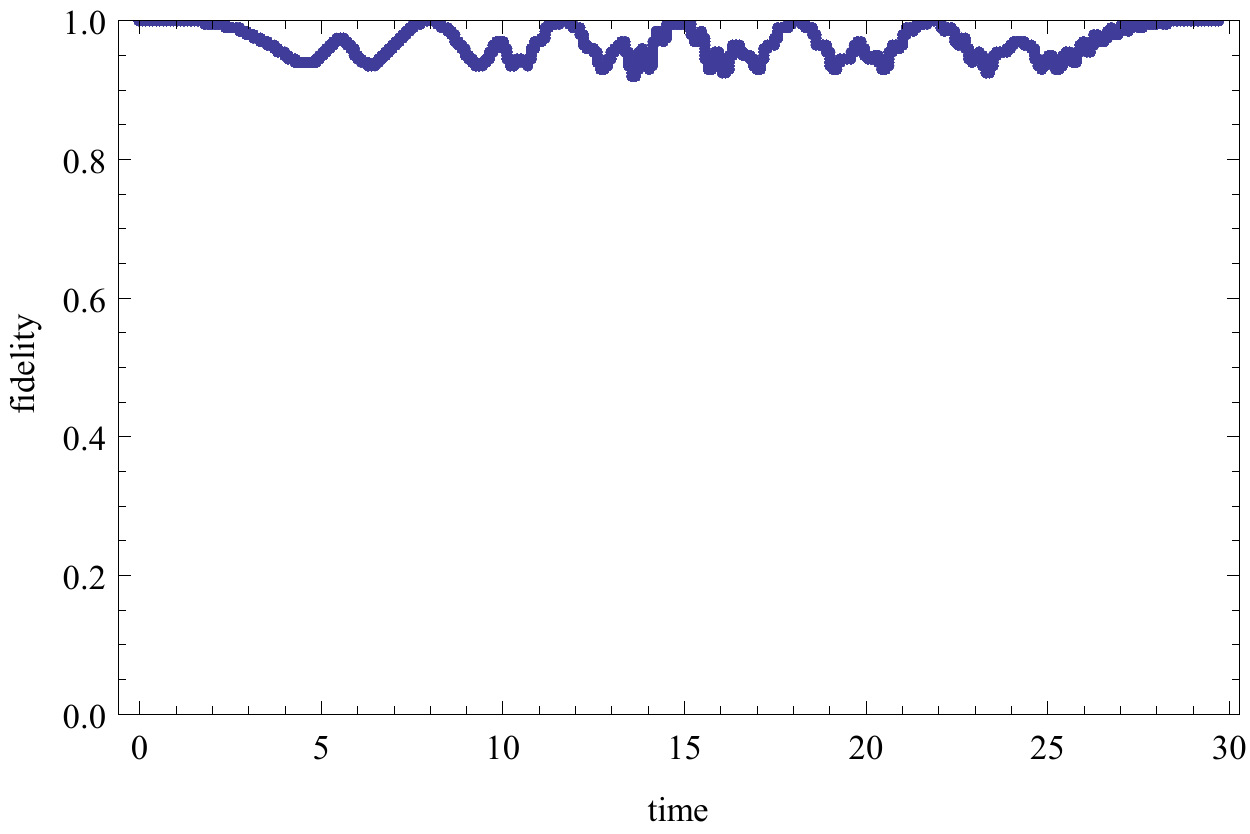}}
\caption{\label{fig:timeevolution}
(Color online) Fidelity as a function of dimensionless time for a spectator atom during the procedure (Step 2)  with $T_{\rm OL} = 1.28$ ms of the two-qubit gate operation.
}
\end{center}
\end{figure}

We also consider the fidelity of spectator atoms to check whether they are left in their ground states of the combined potential $V_{\rm OL} + U_{\rm F}$ during the above described two-qubit gate operation. Figure \ref{fig:timeevolution} shows that the fidelity oscillates with a small amplitude and eventually goes to unity when the atoms undergoing the two-qubit operation are moved back to the initial potential wells. Observe that there are twelve dips in the fidelity corresponding to twelve  peaks of the potential of the moving optical lattice kicking the spectator atom.

\subsection{Step 3}
Let $\psi_0({\bm r})$ be the ground state wave function of an atom in one of the optical lattice potential wells and let $a_s=5.19$~nm be the $s$-wave scattering length between two $^{87}$Rb atoms in $|F=1, m_F=1 \rangle$ and $|F=2, m_F=1\rangle$ \cite{Kaufman}. If two atoms in the ground state are put in the same potential well, the interaction energy is given by
\begin{equation}
\frac{E_{\rm int}}{E_{\rm r}}
 = \frac{2 a_s}{\pi \lambda_{\rm OL}} \int \psi_0^4({\bm r}) d{\bm r} = 0.055.
\end{equation}
The corresponding frequency is $\nu_{\rm int} = 0.0589 \times E_{\rm r} \simeq 218~{\rm Hz}$.
To acquire a phase $\pi$ as necessary for the controlled-$Z$ gate, for example, these atoms must be kept in the same potential well for
\begin{equation}
t_{\rm hold} = \frac{\pi}{2 \pi \nu_{\rm int}}
\simeq 2.29~{\rm ms}.
\end{equation}
This results in a nonlocal two-qubit gate $U = |00\rangle \langle 00| - |01 \rangle \langle 01|+
|10\rangle \langle 10|+|11\rangle \langle 11|$.
The gate $U$ is locally equivalent to the CNOT  and the CZ gates;
application of further one-qubit gates maps  $U$ to these gates.

This step does not involve any nonadiabatic process and we believe the operation
can be executed with a high fidelity by fine tuning the holding time $t_{\rm hold}$.

\subsection{Steps 4 and 5}
After Step 3 is completed, the atoms are returned to their initial positions in the optical lattice (reverse of Step 2) and then the NFFD potentials are switched on (inverse of Step 1).
Since Steps 4 and 5 are the inverse processes of Steps 2 and 1, respectively, the required time and the fidelity of the inverse process are the same as those for a forward process.

\subsection{Execution Time and Fidelity}
We have estimated the time required for each step to give a fidelity of 0.99. Now, let us estimate the overall execution time and  fidelity.
The overall execution time of a two qubit gate is given
by
\begin{equation}
T_{\rm overall} = 2 (T_{\rm F} + T_{\rm OL}) + t_{\rm hold}
\simeq 8.45~{\rm ms}.
\end{equation}
The overall fidelity is estimated as follows. Each of Steps 1, 2, 4 and 5 involves two independent processes. Since these steps involve two atoms, the fidelity associated with each step must be $0.99^2$. Thus, the overall fidelity is $0.99^{8} \sim 0.923$,
where we assumed that the interaction time is tunable so that Step 3 gives a fidelity arbitrarily close to 1.
Compared to our previous proposal, the execution time has not  changed considerably, but the fidelity has been improved since
 the number of steps required for the qubit operation is reduced. The fidelity can be improved further
by spending more time on each step, yielding a longer overall execution time, however.

\section{Summary and discussion}
We have discussed a proposal to demonstrate a selective two-qubit gate operation, which is applied on nearest-neighbor, trapped, neutral atoms. The present setup is less demanding than the previous one proposed in Ref. \cite{Hosseini} because it is no longer necessary to control the size of individual apertures for the NFFD traps. For this new setup, we have obtained an upper bound of the entangled gate operation time $\simeq 8.45~{\rm ms}$ with a corresponding fidelity of $0.923$ bounded by the adiabaticity requirement. Since the two-qubit gate operation consists of a smaller number of processes than that in our previous proposal, the fidelity is expected to be improved.
We believe that this proposal can be demonstrated using current technology, which would be a starting point
toward the realization of a fully-controlled, selective, two-qubit gate operation using neutral atoms.

For the parameters considered in this proposal, with a wavelength of $\simeq 795~{\rm nm}$, an aperture radius of $\simeq 1.2~{\rm \mu m}$ and aperture separation of $\simeq 10~{\rm \mu m}$, fibers with tips of $\simeq 2.4-3~{\rm \mu m}$ should be feasible. For fibers with diameters of $\simeq 3~{\rm \mu m}$, more than $99\%$ of light is confined within the fiber and cross-talk between neighboring fibres, separated by $\simeq 7~{\rm \mu m}$, is negligible \cite{Tong}. In fact, cross-talk between fibers should only be appreciable for subwavelength diameter fibers in which the light is weakly confined and when the fibers are in a strongly coupled configuration (i.e. the fibers are separated by approximately wavelength/10), two conditions which are not relevant in the proposed setup.

\begin{figure}[h]
\begin{center}
{\includegraphics[scale=0.3,clip]{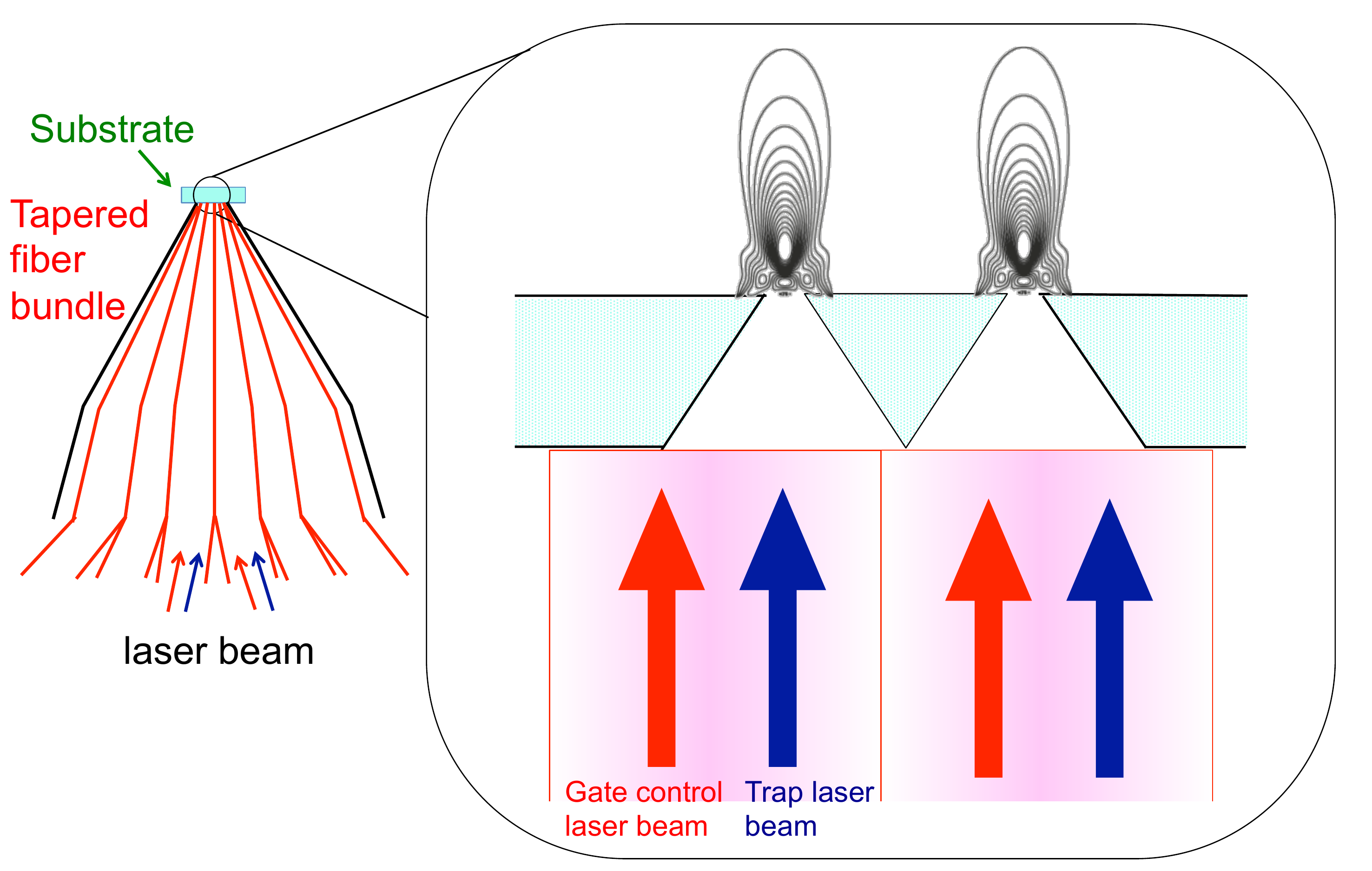}}
\caption{\label{fig:proposal2}
(Color online) Thin apertures shown here may be fabricated by using standard microfabrication methods, wherein the apertures are formed  using an anisotropic wet etching technique \cite{Nijdam}.
The aperture separation is of the order of 10~$\mu$m and, thus, the diameters of individual
fibers in the bundle must be less than the separation. Such a fiber bundle is reported in Ref. \cite{Kosterin}.}
\end{center}
\end{figure}

Our present proposal still maintains scalability in a sense that there are many qubits. If we concentrate only on
 the compatibility check of an NFFD trap \cite{Bandi} and a two-qubit operation \cite{Mandel1,Mandel2}, it is possible to simplify the design
further in order to circumvent difficulties (i) and (ii), as shown in Fig. \ref{fig:proposal2}. This experimental setup is
even less demanding with only two apertures attached to a tapered optical fiber bundle.  We can even remove
the fibers completely, since operations on individual atoms are not necessary.
In this case, only difficulty (iii) remains as a technical challenge.

An alternative to using half-taper fibers would be the integration of optical waveguides, with the aperture array at the end face (output) of the waveguide array. While this would require more sophisticated fabrication techniques, the fragility of the system would be reduced and alignment issues would be eased.

\section*{Acknowledgments}

The work of E. H. L. and T. T. is supported by FIRST project on ``Quantum Information Processing", JSPS, Japan.
Y. K. is partially supported by a Grant-in-Aid for Scientific Research from the Japan Society for the Promotion of Science (No. 25400422).
M. N., and T. O. are partially supported by a Grant-in-Aid for Scientific Research from the Japan Society for the Promotion of Science (No. 23540470).
The authors wish to thank K. Deasy for useful discussions.


\begin{thebibliography}{9}

\bibitem{Nielsen}
M. A. Nielsen and I. L. Chuang: \textit{Quantum computation and Quantum Information}
(Cambridge University Press, Cambridge, U.K., 2000).
%
\bibitem{NakaharaOhmi}
M. Nakahara and T. Ohmi: \textit{Quantum Computing: From Linear Algebra to Physical Realization}
(Taylor and Francis, London, 2008).
%
\bibitem{NatureRev}
T. D. Ladd, F. Jelezko, R. Laflamme, Y. Nakamura, C. Monroe, and J. L. O'Brien,
Nature(London) {\bf 464}, 45 (2010).
%
\bibitem{DiVincenzo}
D. P. DiVincenzo,
Fortschr. Phys. \textbf{48}, 771 (2000).
%
\bibitem{Deutsch}
I. H. Deutsch, G. K. Brennen, and P. S. Jessen,
Fortschr. Phys. \textbf{48}, 925 (2000).
%
\bibitem{Negretti}
A. Negretti, P. Treutlein, and T. Calarco,
Quant. Inf. Process \textbf{10}, 721 (2011).
%
\bibitem{Treutlein}
P. Treutlein, P. Hommelhoff, T. Steinmetz, T. W. H\"{a}nsch, and J. Reichel,
Phys. Rev. Lett. {\bf 92}, 203005 (2004).
%
\bibitem{Schrader}
D. Schrader, I. Dotsenko, M. Khudaverdyan, Y. Miroshnychenko, A. Rauschenbeutel, and D. Meschede,
Phys. Rev. Lett. {\bf 93}, 150501 (2004).
%
\bibitem{Yavuz}
D. D. Yavuz, P. B. Kulatunga, E. Urban, T. A. Johnson, N. Proite, T. Henage, T. G. Walker, and M. Saffman,
Phys. Rev. Lett. \textbf{96}, 063001 (2006).
%
\bibitem{Jones}
M. P. A. Jones, J. Beugnon, A. Ga\"{e}tan, J. Zhang, G. Messin, A. Browaeys, and P. Grangier,
Phys. Rev. A \textbf{75}, 040301(R) (2007).
%
\bibitem{Jaksch}
D. Jaksch, H.-J. Briegel, J. I. Cirac, C. W. Gardiner, and P. Zoller,
Phys. Rev. Lett. {\bf 82}, 1975 (1999).
%
\bibitem{Calarco}
T. Calarco, E. A. Hinds, D. Jaksch, J. Schmiedmayer, J. I. Cirac, and P. Zoller,
Phys. Rev. A {\bf 61}, 022304 (2000).
%
\bibitem{Calarco2}
T. Calarco, U. Dorner, P. S. Julienne, C. J. Williams, and P. Zoller,
Phys. Rev. A {\bf 70}, 012306 (2004).
%
\bibitem{Brennen}
G. K. Brennen, C. M. Caves, P. S. Jessen, and I. H. Deutsch,
Phys. Rev. Lett. {\bf 82}, 1060 (1999).
%
\bibitem{Jaksch2}
D. Jaksch, J. I. Cirac, P. Zoller, S. L. Rolston, R. C\^{o}t\'{e}, and M. D. Lukin,
Phys. Rev. Lett. {\bf 85}, 2208 (2000).
%
\bibitem{Wilk}
T. Wilk, A. Ga\"{e}tan, C. Evellin, J. Wolters, Y. Miroshnychenko, P. Grangier, and A. Browaeys,
Phys. Rev. Lett. {\bf 104}, 010502 (2010).
%
\bibitem{Isenhower}
L. Isenhower, E. Urban, X. L. Zhang, A. T. Gill, T. Henage, T. A. Johnson, T. G. Walker, and M. Saffman,
Phys. Rev. Lett. {\bf 104}, 010503 (2010).
%
\bibitem{Mandel1}
O. Mandel, M. Greiner, A. Widera, T. Rom, T. W. H\"{a}nsch, and I. Bloch,
Phys. Rev. Lett. \textbf{91}, 010407 (2003).
%
\bibitem{Mandel2}
O. Mandel, M. Greiner, A. Widera, T. Rom, T. W. H\"{a}nsch, and I. Bloch,
Nature (London) \textbf{425},  937 (2003).
%
\bibitem{Hosseini}
E. Hosseini Lapasar, K. Kasamatsu, Y. Kondo, M. Nakahara, and T. Ohmi,
J. Phys. Soc. Jpn., \textbf{80}, 114003 (2011).
%
\bibitem{Bandi}
T. N. Bandi, V. G. Minogin, and S. Nic Chormaic,
Phys. Rev. A \textbf{78}, 013410 (2008).
%
\bibitem{Cirac}
O. Romero-Isart, C. Navau, A. Sanchez, P. Zoller, and J. I. Cirac,
Phys. Rev. Lett. \textbf{111}, 145304 (2013).
%
\bibitem{Nakahara}
M. Nakahara, T. Ohmi, and Y. Kondo, arXiv:1009.4426.
%
\bibitem{Bose}
L. Banchi, A. Bayat, P. Verrucchi, and S. Bose,
Phys. Rev. Lett. {\bf 106}, 140501 (2011);
S. Bose, A. Bayat, P. Sodano, L. Banchi, and P. Verrucchi,
\textit{Spin Chains as Data Buses, Logic Buses and Entanglers:
Quantum State Transfer and Network Engineering, Quantum Science and Technology 2014}
(Springer, Heidelberg, 2014) pp. 1-37.
%
\bibitem{Morrissey}
M. Morrissey,K. Deasy, M. Frawley, R. Kumar, E. Prel, L. Russell, V.G. Truong, and S. Nic Chormaic, Sensors \textbf{13}, 10449 (2013).
%
\bibitem{ref:lr}
H. R. Lewis JR. and W. B. Riesenfeld,
J. Math. Phys. {\bf 10}, 1458 (1969).
%
\bibitem{ref:chen1}
E. Torrontegui, S. Ib\'{a}\~{n}ez, X. Chen, A. Ruschhaupt, D. Gu\'{e}ry-Odelin and J. G. Muga,
Phys. Rev. A {\bf 83}, 013415 (2011).
%
\bibitem{ref:chen2}
X. Chen, E. Torrontegui, D. Stefanatos, J.-S. Li, and J. G. Muga,
Phys. Rev. A {\bf 84}, 043415 (2011)
%
\bibitem{Kaufman}
A. M. Kaufman, R. P. Anderson, T. M. Hanna, E. Tiesinga, P. S. Julienne, and D. S. Hall,
Phys. Rev. A {\bf 80}, 050701(R) (2009).
%
\bibitem{Tong}
L. Tong, M. Sumetsky:
\textit{Subwavelength and Nanometer Diameter Optical Fibers},
(Springer, 2010).
%
\bibitem{Nijdam}
A.J. Nijdam, Ph.D. thesis, University of Twente, 2001; http://doc.utwente.nl/38679/.
%
\bibitem{Kosterin}
A. Kosterin, V. Temyanko, M. Fallahi, and M. Mansuripur,
Applied Optics {\bf 43}, 3893 (2004).

\end{thebibliography}
\end{document}